\documentclass[aps,prb,twocolumn,amsmath,amssymb,superscriptaddress]{revtex4}
\usepackage[dvips]{graphicx}
\usepackage{dcolumn}
\usepackage{bm}
\usepackage{float,color}
\usepackage{gensymb}
\usepackage{threeparttable}
\usepackage{appendix}
\usepackage{rotfloat}
\usepackage{array}
\newcommand{\PreserveBackslash}[1]{\let\temp=\\#1\let\\=\temp}
\newcolumntype{C}[1]{>{\PreserveBackslash\centering}p{#1}}
\newcolumntype{R}[1]{>{\PreserveBackslash\raggedleft}p{#1}}
\newcolumntype{L}[1]{>{\PreserveBackslash\raggedright}p{#1}}

\begin{document}
\title{
High temperature superconducting phase of HBr 
under pressure predicted by first-principles calculations}

\author{Qinyan Gu}
\affiliation{
National Laboratory of Solid State Microstructures,
School of Physics and Collaborative Innovation Center of Advanced Microstructures,
Nanjing University, Nanjing, 210093, P. R. China}

\author{Pengchao Lu}
\affiliation{
National Laboratory of Solid State Microstructures,
School of Physics and Collaborative Innovation Center of Advanced Microstructures,
Nanjing University, Nanjing, 210093, P. R. China}

\author{Kang Xia}
\affiliation{
National Laboratory of Solid State Microstructures,
School of Physics and Collaborative Innovation Center of Advanced Microstructures,
Nanjing University, Nanjing, 210093, P. R. China}

\author{Jian Sun}
\email[Correspondence should be addressed to J.S. ]{(E-mail: jiansun@nju.edu.cn)}
\affiliation{
National Laboratory of Solid State Microstructures,
School of Physics and Collaborative Innovation Center of Advanced Microstructures,
Nanjing University, Nanjing, 210093, P. R. China}

\author{Dingyu Xing}
\affiliation{
National Laboratory of Solid State Microstructures,
School of Physics and Collaborative Innovation Center of Advanced Microstructures,
Nanjing University, Nanjing, 210093, P. R. China}

\date{\today}

\begin{abstract}

The high pressure phases of HBr are explored 
with an $ab$ $initio$ crystal structure search. 
By taking into account the contribution of zero-point energy (ZPE),
we find that the $P4/nmm$ phase of HBr 
is thermodynamically stable in the pressure range from 150 to 200 GPa.
The superconducting critical temperature ($T_c$) of $P4/nmm$ HBr 
is evaluated to be around 73 K at 170 GPa, 
which is the highest record so far among binary halogen hydrides. 
Its $T_c$ can be further raised to around 95K under 170 GPa 
if half of the bromine atoms in the $P4/nmm$ HBr 
are substituted by the lighter chlorine atoms. 
Our study shows that, in addition to lower mass, 
higher coordination number, 
shorter bonds, and more highly symmetric environment 
for the hydrogen atoms are important factors 
to enhance the superconductivity in hydrides.
\end{abstract}

\maketitle

\section{Introduction}

In 1935, Wigner and Huntington envisioned that, 
at low temperature, hydrogen molecular 
would transform into atomic metallic solid hydrogen under 25 GPa \cite{Wigner1935}. 
Later Ashcroft \cite{Ashcroft1968} suggested that 
hydrogen would become a room temperature superconductor under high pressure. 
After that, tremendous efforts have been invested into the research 
on the synthesis of metallic hydrogen and its properties, 
both in theory and experiments 
\cite{Lorenzana1989, Loubeyre2002, Pickard2007, Cudazzo2008, 
McMahon2011, Eremets2011, Howie2012, Magdau2013, McWilliams2016, Dalladay-Simpson2016, Huang2016, Dias2017}. 
The mechanism of the superconductivity in metallic hydrogen 
is believed to be phonon mediated 
and will be greatly enhanced under high pressure. 
For instance, the metallic hydrogen in $Cmca$ structure was predicted to 
have a $T_c$ of 242 K near 450 GPa \cite{Cudazzo2008}
and would reach higher $T_c$ 
when hydrogen is compressed into denser structures \cite{McMahon2011}. 
On the experimental side
\cite{Lorenzana1989, Eremets2011, Howie2012, Magdau2013, McWilliams2016, Dalladay-Simpson2016, Huang2016,Dias2017},  
the experimental observation of the metallic hydrogen is still under debate,
and the realization of the superconductivity in solid hydrogen 
seems to be far away.

Due to the reason that 
the critical pressure of the metallization of hydrogen is high, 
it will be difficult to put it into wide applications. 
Ashcroft \cite{Ashcroft2004} proposed to metallize hydrogen 
together with a monatomic or paired metal
by using a method of chemical precompression. 
Since then, compared to hydrogen,
a large number of hydrides 
with superconducting properties at lower pressure 
have been proposed theoretically
and experimentally studied. 
SiH$_4$ \cite{Ashcroft2004}, as a member of IVA\_hydrides, 
was predicted to have equivalent electron density 
compared with pure hydrogen theoretically when compressed to 100 GPa. 
This compound became the first example of superconducting hydrides in experiment, 
with a $T_c$ of around 17 K at 96 and 120 GPa \cite{Eremets2008}.
Then, quite a few hydrogen-rich compounds were predicted to
have high $T_c$ values based on $ab$ $initio$ calculations:
SiH$_3$ with $T_c$ \~{} 139 K at 275 GPa \cite{Jin2010},
CaH$_6$ with $T_c$ of about 220-235 K at 150 GPa \cite{Wang2012-CaH}
and so on \cite{Tse2007, Gao2008, Flores-Livas2012}.

Recently, the sulfur hydrogen system has been identified experimentally 
to be remarkable high-temperature superconductivity 
($T_c$ up to 203 K) under pressure \cite{Drozdov2015}. 
The measured $T_c$ of H$_3$S is consistent with
the theoretical results \cite{Duan2014}
based on Bardeen-Cooper-Schrieﬀer (BCS) theory \cite{Allen1975}.
This further strengthens Ashcroft's opinion 
that the metallization and superconductivity could be realized 
in hydrogen-dominant compounds \cite{Ashcroft2004}. 
Moreover, the $T_c$ has been predicted to be enhanced 
by increasing the ionic character in the H$_3$S compound 
with hypothetical alchemical mixture of oxygen \cite{Heil2015} 
and phosphorus \cite{Ge2016} atoms. 
Where the $T_c$ of H$_3$S$_{0.925}$P$_{0.075}$ 
in the $Im-$3$m$ structure at 250 GPa 
was predicted to be around 280 K \cite{Ge2016}. 
Very recently, 
as an analog of H$_2$S, 
PH$_3$ was reported to possess  
a $T_c$ of about 100 K at 200 GPa in experiments \cite{Drozdov2015}.

Among the hydrides, 
the exploration of the superconductivity in halogen hydrides 
is impelled by 
the metallization of HCl and HBr \cite{Zhang2010}, 
where the hydrogen-bond symmetrization 
plays a critical role 
\cite{Sequeira1972, Chen1997, Duan2010, Zeng2017}. 
The $T_c$ of $Cmcm$-H$_2$Br and $P$6$/mmm$-H$_4$I 
were theoretically predicted to be around 12.1 K at 240 GPa \cite{Duan2015a} 
and 17.5 K at 100 GPa \cite{Shamp2015}, respectively. 
HCl in $C$2$/m$\_symmetry was calculated 
to have a $T_c$ of 20 K at 250 GPa \cite{Chen2015}. 
It seems that the $T_c$ values of H-rich halogen hydrides 
are not very high compared to other hydrides.

In a quantum mechanical system,
although the static lattice energy usually contributes the largest part to the total energy,
the ZPE can be crucial for determining the stability 
especially when the energy difference between phases is small.
The influence of the ZPE on the energetic stability
has been comprehensively tested 
in the studies of the phase diagram of hydrogen
\cite{Natoli1993, McMahon2011, Azadi2014, Drummond2015}.
Particularly, ZPE tends to be larger in the high symmetric structure,
compared to the low symmetric structure
\cite{Straus1977, Ceperley1987}.
In a previous work of HBr system \cite{Chen2015}, 
neglecting the effect of ZPE, a $C2/m$ structure is found 
to be the thermodynamically stable one rather than the $P4/nmm$ structure.
However, due to the facts that 
HBr is a system containing light elements and
the $P4/nmm$ structure has a higher symmetry than the $C2/m$ structure,  
we believe it is necessary 
to consider the effect of ZPE in the HBr system.

In this paper, 
with additional consideration of ZPE, 
the $P4/nmm$ HBr is found to be more stable than the $C2/m$ phase above 150 GPa. 
More interestingly, 
the estimated value of the superconducting transition temperature ($T_c$) of the $P4/nmm$\_HBr 
can get a high value of 73 K (at 170 GPa). 
Furthermore,
we find that an atomic substituted structure of HCl$_{0.5}$Br$_{0.5}$ 
can realize an even higher $T_c$ of around 95 K at 170 GPa. 
In addition to the significant effects of low atomic mass,  
here we reveal that 
high coordination number, shorter bonds, and a more symmetrically restricted environment 
of hydrogen atoms play key roles in enhancing superconductivity.

\section{Methods}

We used an $ab$ $initio$ random structure searching method (AIRSS)
\cite{Pickard2006, Pickard2011}
to investigate the stable/metastable phases of HBr 
under high pressure up to 200 GPa and different cell-size
up to 18 atoms. 
%
%
Structural optimization and electronic structures calculations 
were carried out by projector augmented wave method 
implemented in the Vienna $ab$ $initio$ simulation package (VASP) \cite{Kresse1996}.
Perdew-Burke-Ernzerhof (PBE) \cite{Perdew1996}
and the PBE functional revised for solids (PBEsol) \cite{Perdew2008} 
exchange and correlation 
were employed to investigate the structural stability of HBr,
together with Grimme’s DFT-D2 method \cite{Grimme2006}
for the van der Waals (vdW) interaction.
The plane-wave kinetic cutoff was 400 eV, 
and the Brillouin zones were sampled 
by Monkhorst-Pack method \cite{Monkhorst1976} 
with a k\_spacing of $0.02\times{2\pi}$\AA$^{-1}$. 
ZPE was obtained from the quasiharmonic approximation \cite{Cazorla2013, Cazorla2015}
from phonon calculations, by 
finite displacement method implemented in PHONOPY code \cite{Togo2008},
 with a $3\times3\times4$ and $3\times3\times2$ supercell for HBr 
with $P4/nmm$ and $C$2$/m$ phase, respectively.
Electron-phonon coupling (EPC) calculations were executed 
in the framework of density functional perturbation theory,  
as implemented in the QUANTUM-ESPRESSO code \cite{Giannozzi2009}. 
We adopted a $16\times16\times20$ k-point mesh 
for charge self-consistent calculation, 
a $32\times32\times40$ k-point mesh for EPC linewidth integration 
and a $4\times4\times5$ q-point mesh for dynamical matrix. 
Norm-conserving pseudopotentials were used 
with the energy cutoffs of 160 Ry 
for the wave functions and 640 Ry 
for the charge density to ensure  
that the error of total energy was less than 10$^{-6}$ Ry.
 
\section{Results and discussions}

\subsection{Crystal Structure and phase diagram}

\begin{figure}[hpt]
\begin{center}
\includegraphics[width=0.45\textwidth]{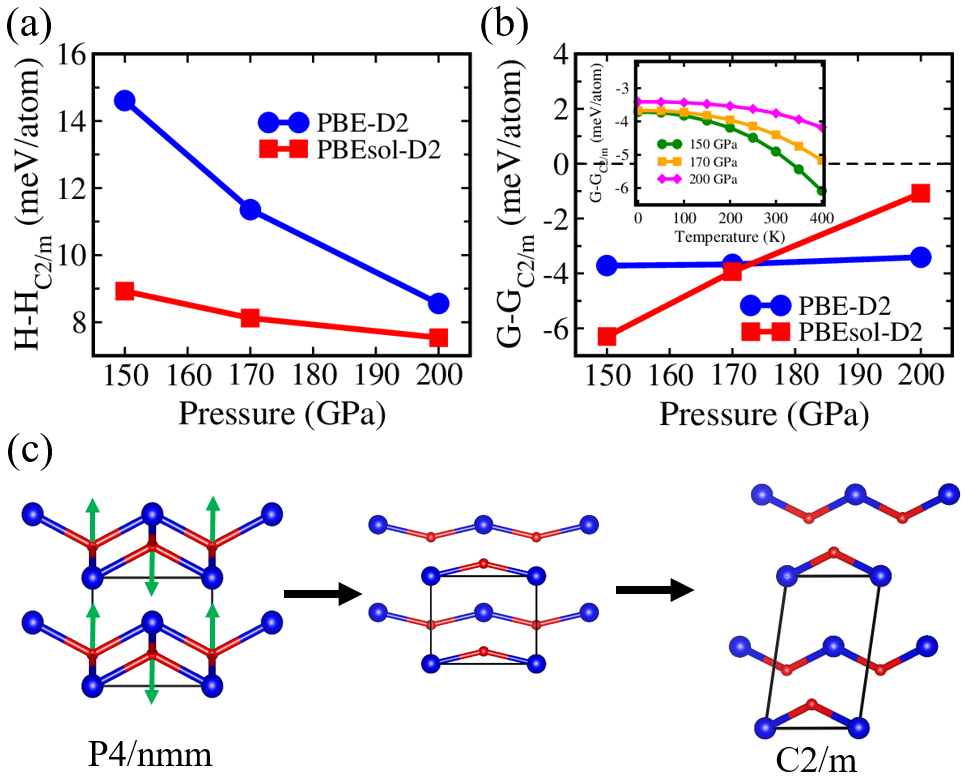}
\caption{\textcolor{blue}{(Color online)}
The calculated enthalpy of $P4/nmm$ HBr 
relative to the $C2/m$ HBr under pressure, 
given by the PBE-D2 and PBEsol-D2 functionals, 
(a) without and (b) with considering the ZPE, respectively. 
The inset shows the difference of Gibbs free energy of $P4/nmm$ 
HBr relative to $C2/m$ HBr at different 
temperature and pressure, 
with PBE-D2 functional. 
(c) From $P4/nmm$ HBr to $C2/m$ phase,
 with bond breaking and interlayer sliding. 
Once the hydrogen atom moves away from 
the center of the HBr$_4$ tetrahedron, 
half of the H-Br bonds in the $P4/nmm$ phase will be broken, 
which leads to forming a 1D chain-like structure.
}
\label{fig:enthalpy}
\end{center}
\end{figure}

Neglecting the effect of ZPE,
three energy-favorable phases are identified 
during our crystal structure search. 
Two of them are stable in enthalpy: 
the $P-1$ phase (space group No. 2) below 120 GPa, 
and the $C2/m$ phase (space group No. 12) above 120 GPa, 
while the other one, the $P4/nmm$ phase (space group No. 129) 
is metastable, whose enthalpy is slightly highter than that   
of the $C2/m$ phase (around 10 meV/atom). 
The calculated enthalpies are shown in Fig.\ \ref{fig:enthalpy}(a).
The results given by the two different exchange-correlation functionals 
show the same tendency.
In addition, the functional PEBsol-D2 
reduces the enthalpy difference between 
$P4/nmm$ and $C2/m$ phase, compared with PBE functional.  
Both the structures and transition pressures 
are consistent with the previous study \cite{Chen2015}.

As mentioned above, 
ZPE is sometimes crucial to judge the stability of structures 
for compounds with light elements \cite{Errea2016};
it has a more notable effect 
on high-symmetry structures compared with low-symmetry structures. 
Due to the fact that $C2/m$ and $P4/nmm$ 
structures have very different symmetry, 
ZPE might change their stability sequence. 
Our calculations [presented in Fig.\ \ref{fig:enthalpy}(b)] 
show that the $P4/nmm$ phase indeed 
has lower Gibbs free energy than 
$C2/m$ phase under high pressure. 
The Gibbs free energy difference between 
$C2/m$ and $P4/nmm$ phase becomes larger with the increasing of temperature. 
Contrary to the earlier report \cite{Chen2015},
our calculations suggest that 
the thermodynamically stable phase of HBr 
under pressure above 150 GPa is the $P4/nmm$ phase, 
instead of the $C2/m$ phase.
The hydrogen and bromine atoms in the $C2/m$ structure
are all two-coordinated 
and they assemble into 1D chain-like structure,
while the hydrogen and bromine atoms in the $P4/nmm$ structure 
are all fore-coordinated and they form a 2D netlike structure. 
Actually, the $P4/nmm$ and $C2/m$
fulfill a group-subgroup relationship, 
$C2/m$ is a subgroup of $P4/nmm$ with an index of 4.
If we break half of the H-Br bonds in the H-centered tetrahedrons 
in the $P4/nmm$ structure, 
the 2D net will break into 1D chains. 
Then with slight distortion and interlayer sliding, 
the structure will easily transform into the $C$2$/m$ structure. 
This might be the reason 
why their enthalpy difference 
at high pressure is so small.
Another point that needs to be mentioned is that, 
due to the 2D and 1D features in the $P4/nmm$ and $C2/m$ structures, 
and that vdW corrections are crucial to describe 2D and molecular systems 
\cite{Lu2016, Xia2017},
the DFT-D2 type of vdW correction \cite{Grimme2006} is used in this work. 

Phonon spectra is often used to verify 
the dynamical stability of structures. 
In this case, there is no imaginary frequency for the $P4/nmm$ HBr 
upon the pressure above 150 GPa,
indicating its dynamical stability. 
Both hydrogen and bromine atoms are monovalent;
consequently, they tend to form diatomic molecule 
and hydrogen bonds between these diatomic molecules. 
High pressure enhances the atomic orbital overlaps and 
pushes the molecules to polymerize with each other 
to form chains or squares, 
and then further compression leads to symmetrization of the hydrogen bonds. 
However, in most of the symmetric hydrogen bond cases, 
the hydrogen atom is two-coordinated,
while in the $P4/nmm$ structure, 
the hydrogen atoms are four-coordinated; 
that is probably why it can only appear 
at such a high pressure above 150 GPa. 

\subsection{Electronic structures}

\begin{figure}[hpt]
\begin{center}
\includegraphics[width=0.45\textwidth]{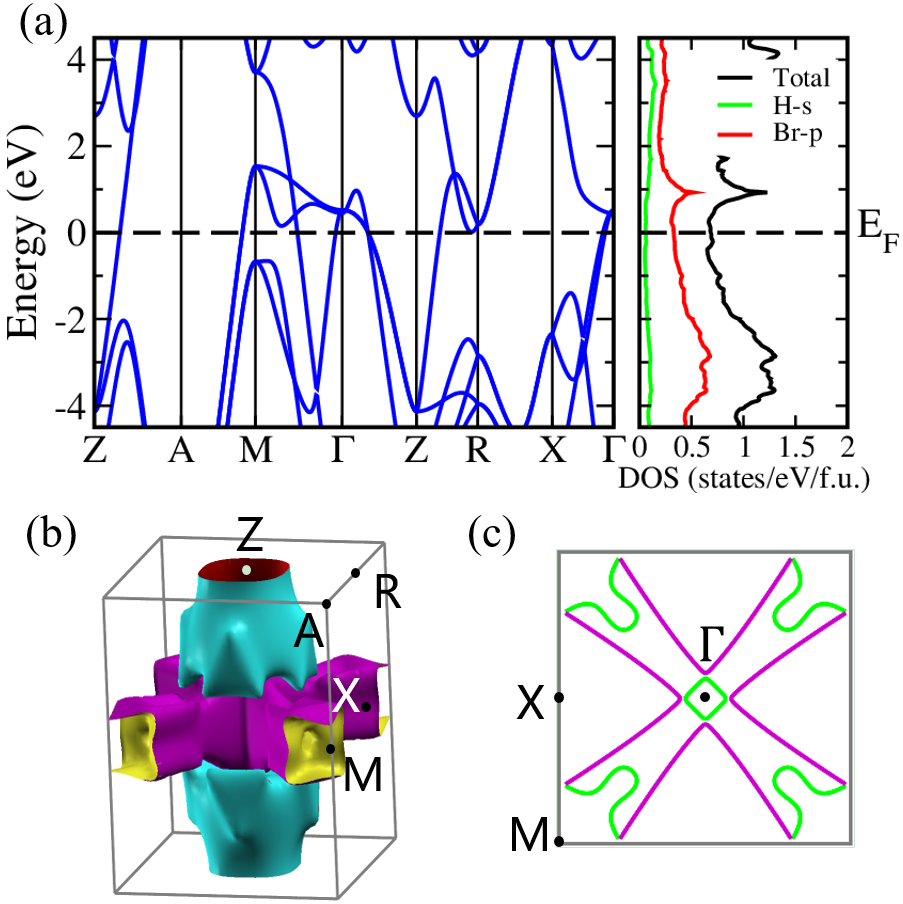}
\caption
{\textcolor{blue}{(Color online)}
(a) Calculated band structures, 
projected density of states (DOS), 
and (b) Fermi surface of $P4/nmm$ HBr at 170 GPa.
(c) The contour plot of Fermi surface 
at the plane of $k_z$ = 0.
}
\label{fig:band1}
\end{center}
\end{figure}

To explore the electronic properties of 
the $P4/nmm$ phase of HBr under pressure, 
we calculate the electronic band structure, 
projected density of states (DOS) and Fermi surface.
 As shown in Fig.\ \ref{fig:band1}(a), 
the existence of large dispersion bands crossing 
the Fermi level indicates the metallic character. 
It reveals that states near the Fermi level are mainly 
composed by the $p$ orbital of bromine atoms. 
There are three components forming the Fermi surface: 
a bonelike electron pocket around the $Z$ point, 
and two nesting hole pockets crossing the $k_z$ = 0 plane 
and composed of orthogonal parallel tubes 
[presented in Fig.\ \ref{fig:band1}(b)]. 
The Fermi nesting 
from the parallel Fermi surfaces 
nearly along the $q_M$ (0.5, 0.5, 0.0) direction
can be seen clearly, 
from both Figs.\ \ref{fig:band1}(b) and \ \ref{fig:band1}(c).
This nesting feature of the Fermi surface  
resulting from the tetrahedral symmetry could be 
beneficial to the superconductivity.

\begin{figure}[hpt]
\begin{center}
\includegraphics[width=0.45\textwidth]{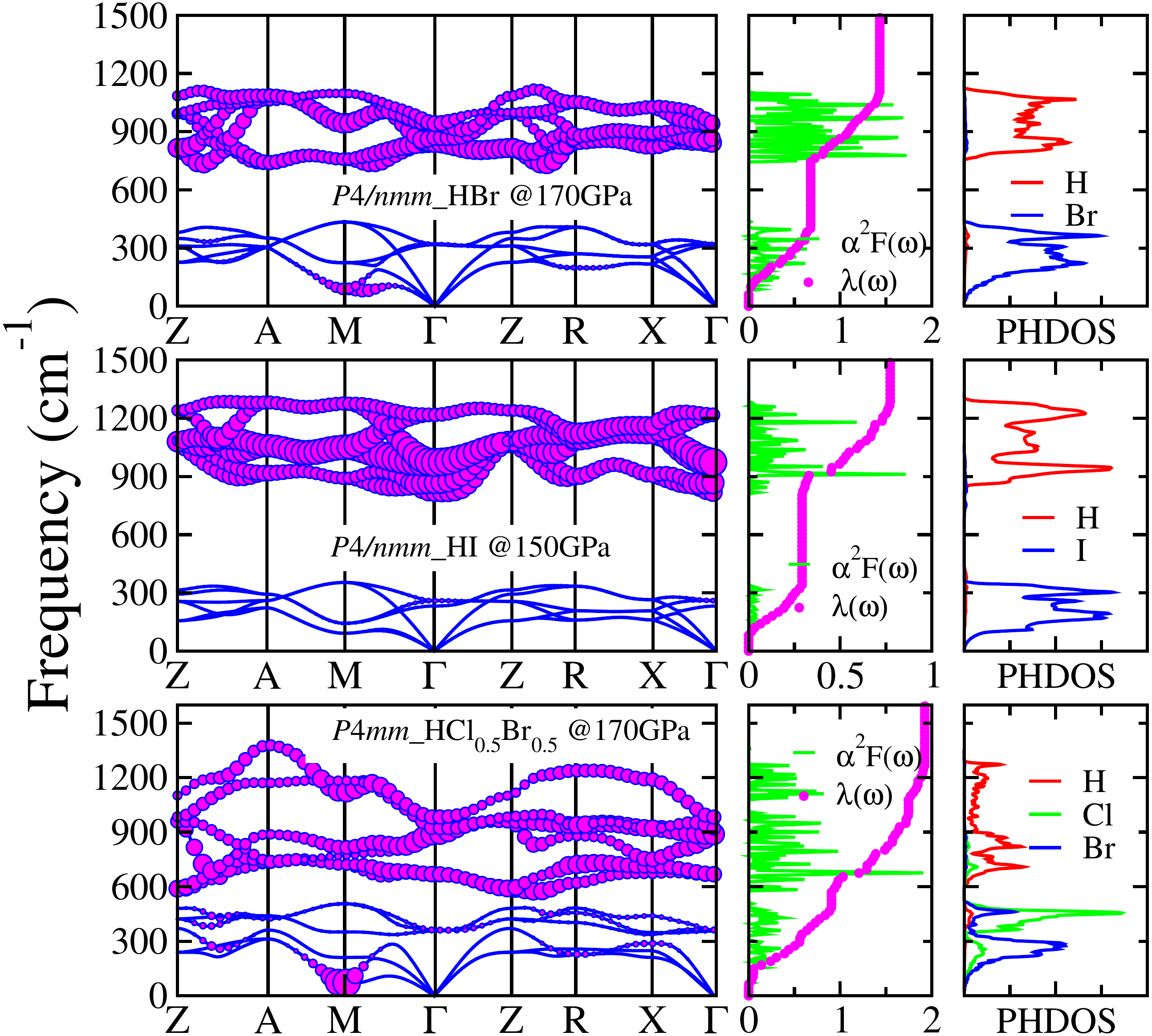}
\caption{\textcolor{blue}{(Color online)}
Phonon dispersion curves, Eliashberg spectral functions $\alpha^2F$($\omega$) 
together with the electron-phonon integral $\lambda$($\omega$) 
and phonon density of states (PHDOS) for  $P4/nmm$-HBr,
 $P4mm$-HCl$_{0.5}$Br$_{0.5}$ at 170 GPa, and  $P4/nmm$-HI at 150 GPa.
 The phonon linewidth $\gamma_{q,j}$($\omega$) of each mode (q,j) 
is illustrated by the size of pink circles along the phonon dispersions. 
}
\label{fig:elph}
\end{center}
\end{figure}

\subsection{Atomic substitution vs. Electron-phonon coupling}

\begin{figure}[hpt]
\begin{center}
\includegraphics[width=0.45\textwidth]{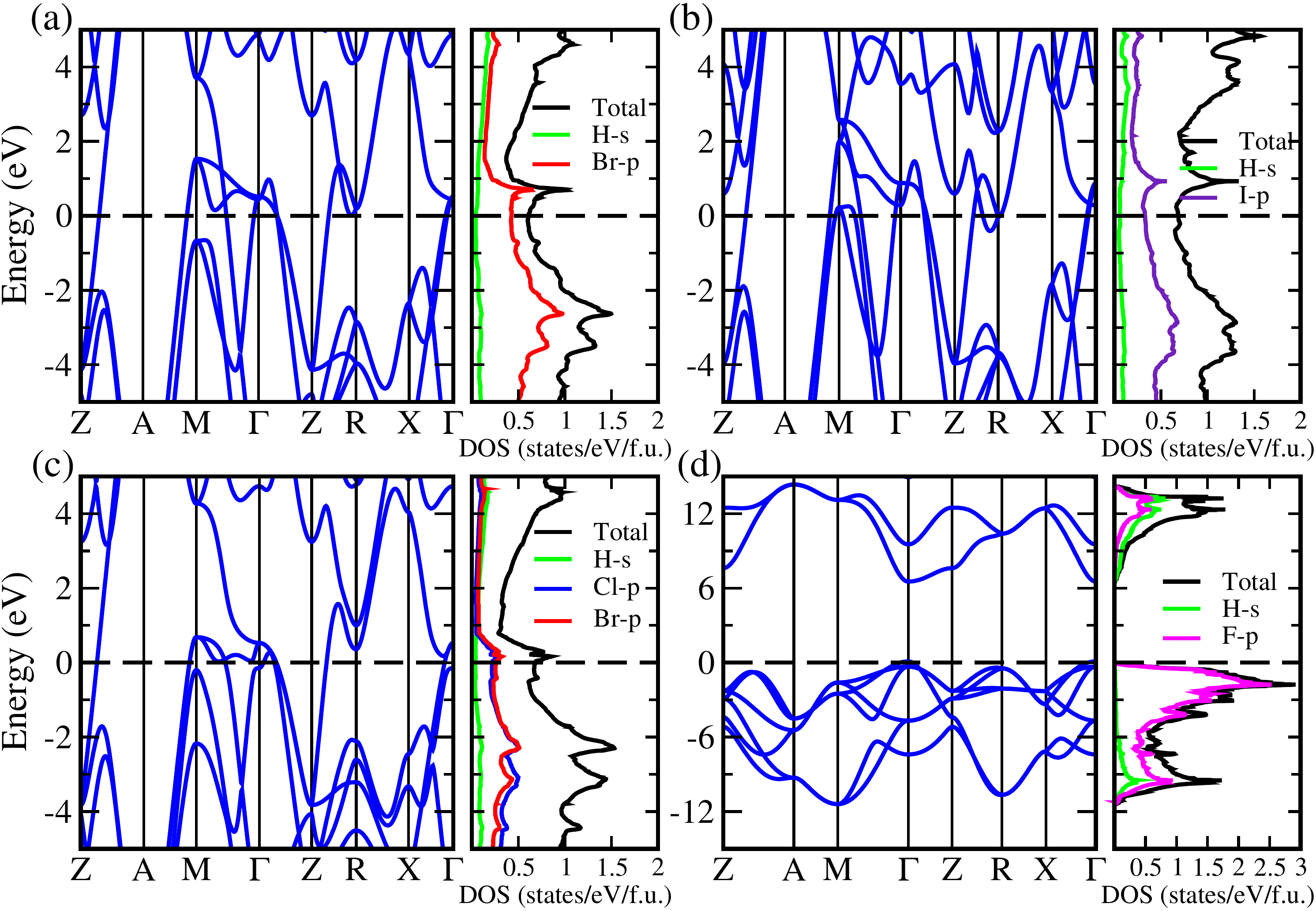}
\caption{\textcolor{blue}{(Color online)}
The calculated band structures and 
density of states (DOS) for the $P4/nmm$ structure of 
(a) HBr at 170 GPa, (b) HI at 150 GPa and (d) HF at 150 GPa.
(c) DOS for the $P4mm$ structure of HCl$_{0.5}$Br$_{0.5}$ at 170 GPa.
}
\label{fig:band2}
\end{center}
\end{figure}

As atomic substitution is a common way to induce or enhance 
superconductivity properties of materials in experiments 
\cite{Cava1990, Ekimov2004, Kimber2009}, 
we try to raise the 
$T_c$ in this system 
with replacing bromine atoms by other lighter halogen atoms. 
In our calculations, 
the $P4/nmm$-HF exhibits insulating character, 
as shown in Fig.\ \ref{fig:band2}(d). 
The stable pressure for $P4/nmm$ HCl is higher than 200GPa, 
which agrees with the previous studies \cite{Chen2015, Zeng2017}. 
The $P4/nmm$ HBr is stable in the pressure from 150 to 200GPa.
Therefore, a possible way to raise $T_c$
is to introduce half substitution of bromine by chlorine atoms. 
The symmetry of the resulted HCl$_{0.5}$Br$_{0.5}$ compound 
then becomes to $P4mm$.
To illustrate the thermodynamical stability of the $P4mm$ HCl$_{0.5}$Br$_{0.5}$
compound, we calculate its forming enthalpy relative to H$_2$+Cl$_2$+Br$_2$, 
using the stable phase $C2/c$ of the solid H$_2$ \cite{Pickard2007},
phase $Immm$ of the solid Cl$_2$ \cite{Li2012} 
and phase $Cmca$ of solid Br$_2$ \cite{Powell2006}
at the pressure from 150 to 200 GPa. 
From Table.\ \ref{table:HClBrenthalpy}, 
it can be clearly seen that the $P4mm$\_HCl$_{0.5}$Br$_{0.5}$
is fairly stable against decomposition.

We performed EPC calculations
for HBr, HI in $P4/nmm$ structure and
HCl$_{0.5}$Br$_{0.5}$ in $P4mm$ structure,
presented in Fig.\ \ref{fig:elph}.
Here, the Allen-Dynes modified McMillian equation \cite{Allen1975},
\begin{displaymath}\label{qq}
T_c=\frac{\omega_{log}}{1.2} \exp{[{-}\frac{1.04 (1+\lambda)}{(\lambda-\mu^*(1+0.62\lambda))}]}
\end{displaymath}
is used to probe their potential superconductivity. 
From Fig.\ \ref{fig:elph}, the low-frequency vibrations
are related to the halogen atoms and the high-frequency modes
come from the vibrations of the hydrogen atom.
The H-stretching modes in HBr give prominent contribution
(about 59\%) to the integral EPC constant $\lambda$.
From Table.\ \ref{table:$T_c$}, 
the results shows that 
the integral EPC parameter $\lambda$ changes from $1.43$ in HBr to $1.93$
in HCl$_{0.5}$Br$_{0.5}$ at 170 GPa.
With a common Coulomb screening constant value of $\mu^*$ = 0.1, 
the $T_c$ of HBr is estimated to be around 73K and
HCl$_{0.5}$Br$_{0.5}$ can reach around 95 K at 170 GPa,
while the $T_c$ of $P4/nmm$ HI is only about 47 K at 150 GPa. 

\begin{figure}[hpt]
\begin{center}
\includegraphics[width=0.45\textwidth]{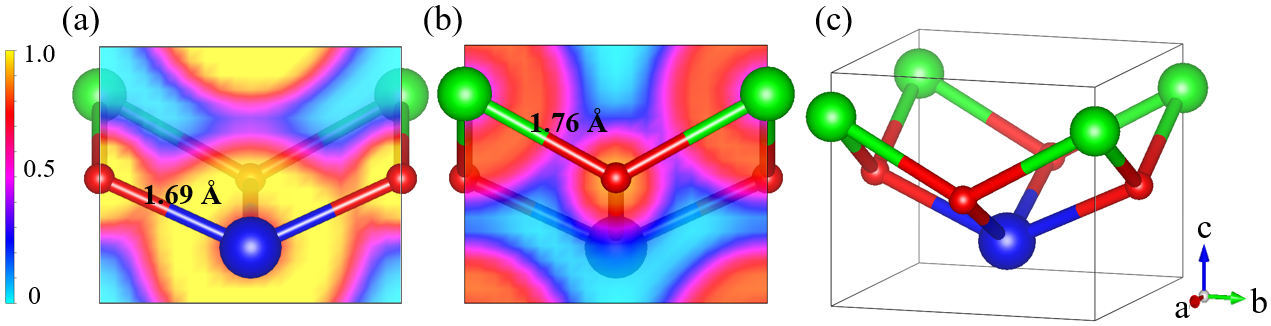}
\caption{\textcolor{blue}{(Color online)}
Contour plots of electron localization function (ELF)
on the (100) plane along the (a) H-Br and (b) H-Cl bonds, respectively,
in the $P4mm$ HCl$_{0.5}$Br$_{0.5}$ at 170 GPa.
(c) Crystal structure of $P4mm$ HCl$_{0.5}$Br$_{0.5}$.
The red, green, and blue balls represent
the hydrogen, chlorine, and bromine atoms, respectively. 
}
\label{fig:elf}
\end{center}
\end{figure}

To get more insights on the reason why the half substitution on HBr 
can increase $T_c$, 
the electronic structures of halogen hydrides 
are calculated in Fig.\ \ref{fig:band2}. 
We can see that the DOS near the Fermi level in HCl$_{0.5}$Br$_{0.5}$
increases compared to that in the HBr and HI compounds and 
the Van Hove singularity in HCl$_{0.5}$Br$_{0.5}$ is closer to the Fermi level,
which is mainly resulting from the halogen atoms.
This increase of DOS near the Fermi level may
enhance the value of $T_c$ \cite{Ge2016}. 
Meanwhile, we analyze the chemical environment of hydrogen atoms 
in halogen hydrides by calculating the bond length and Bader charges, 
which are summarized in Table.\ \ref{table:bader charge}. 
The chlorine, bromine, and iodine atoms have the same number of valence electrons, 
and the chlorine atom has the strongest electronegativity and 
least radius among them. 
It is clear that the length of H-I bondis longer than the H-Br bond 
in the $P4/nmm$ structure. 
However, in the $P4mm$ structure of HCl$_{0.5}$Br$_{0.5}$, 
it is abnormal that the length of H-Cl bond is longer than the H-Br bond.
Here, we calculate the ELF of $P4mm$ structure for HCl$_{0.5}$Br$_{0.5}$ at 170 GPa 
(shown in Fig.\ \ref{fig:elf}) to explain it. 
From the Bader charge analysis, 
the bromine atom changes from the anion to the cation 
because of the strong electronegativity of substitution of the chlorine atom.
So the electrons prefer to locate between the hydrogen and bromine atoms
and repel the chlorine atoms, 
which causes the abnormal elongation of the H-Cl bond and the shortening of the H-Br bond.The mid-lying optic phonon bands in HCl$_{0.5}$Br$_{0.5}$
are somehow softened 
due to the unexpected weak H-Cl bond in $P4mm$ HCl$_{0.5}$Br$_{0.5}$.
The shorter H-Br bonds 
make the highest phonon frequency harder
compared with that in $P4/nmm$ HBr.
Meanwhile, we can see that the gap between the low and high frequencies
becomes smaller with the substitution.
The increase of intermediated phonon linewidth will enhance
the overall electron-phonon coupling constant.

\begin{center}
\begin{table}
  \caption{
        The calculated Bader charge and bond length of HBr, HI, HCl$_{0.5}$Br$_{0.5}$, and HF. The negative number in \textit{$\Delta$} charge represents the number of missing electrons.
}
 \begin{tabular}[t]{m{2.8cm}m{1.3cm}m{1.3cm}p{2.3cm}}
        \hline\hline
        Compound &   atom  & \textit{$\Delta$}charge &   Bond length (\AA) \\
        \hline
        $P4/nmm$ HBr & H  & -0.03   &  H-Br = 1.77 \\
        (170 GPa)    & Br &  0.03 & \\
        \hline
        $P4/nmm$ HI  & H  &  0.23  &   H-I = 1.93 \\
        (150 GPa)    & I  & -0.23 \\
        \hline
        $P4mm$     & H  & -0.14   &  H-Cl = 1.76 \\
        HCl$_{0.5}$Br$_{0.5}$    & Cl &  0.40    &  H-Br = 1.69 \\
        (170 GPa)             & Br & -0.26 \\
        \hline
        $P4/nmm$ HF  & H &  -0.64   &  H-F = 1.34  \\
        (150GPa)     & F &   0.64 &\\
        \hline\hline
  \end{tabular}
\label{table:bader charge}
\end{table}
\end{center}

\begin{figure}[hpt]
\begin{center}
\includegraphics[width=0.45\textwidth]{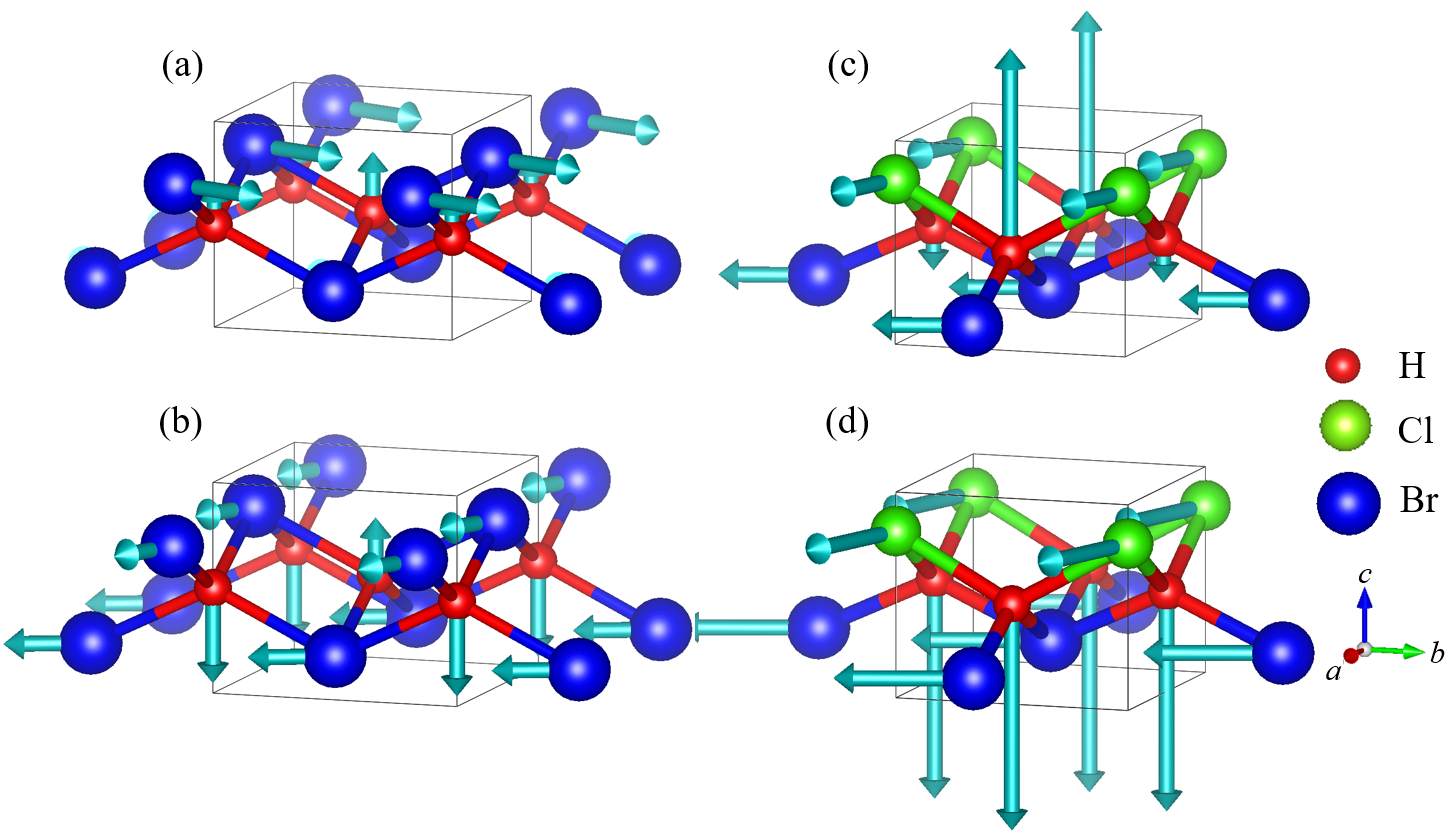}
\caption{\textcolor{blue}{(Color online)}
The vibrational modes at the $M$ point
for the lowest twofold degenerate frequency
of (a), (b) $P4/nmm$ HBr
and (c), (d) $P4mm$ HCl$_{0.5}$Br$_{0.5}$.
They belong to a 2D $M_3$ irreducible representation.
}
\label{fig:mode}
\end{center}
\end{figure}

We note that the low-energy modes at M point goes softening;
for both HBr and HCl$_{0.5}$Br$_{0.5}$,
the large phonon linewidth of these modes indicates that 
they are strongly coupled to the electrons.
In these modes (see Fig.\ \ref{fig:mode}),
the vibrations of halogen atoms are
in the $ab$ plane with similar strength both in HBr and HCl$_{0.5}$Br$_{0.5}$,
while the vibrations of hydrogen atoms are
perpendicular to the $ab$ plane.
The hydrogen vibrations in HCl$_{0.5}$Br$_{0.5}$
are much stronger than that in HBr,
which makes the Br-H-Br and Cl-H-Cl bending amplitude
much larger than that of the Br-H-Br bending in HBr.
Large amplitude collective motions are usually in low frequency \cite{Sun2014}.
Although they might be renormalized
by anharmonicity for the low frequency,
it seems that they do not have a large contribution
to the integral EPC constant $\lambda$ as one can see from Fig.\ \ref{fig:elph}.

\subsection{$T_c$ and symmetric environment for hydrogen}

It was reported earlier that the $T_c$ of $C2/m$ HBr is
around  $9.7\times10^{-3}$ K at 120 GPa \cite{Chen2015},
and  HBr in the $P4/nmm$ structure exhibits
a much superior superconducting transition temperature (about 73 K at 170 GPa).
The dramatic difference in superconducting $T_c$
between these two phases indicates that the environment around hydrogen atoms
plays a significant role in superconductivity.
As we mentioned above,
each hydrogen atom in the $P4/nmm$ structure 
has four tetrahedrally arranged H-Br bonds.
This symmetrical environment around hydrogen atoms would
harden the H-stretching modes by holding back the vibration of hydrogen atoms.
Meanwhile, the average H-Br distance changes from 1.87 \AA\ in $C2/m$
to 1.79 \AA\ in $P4/nmm$ at 150 GPa.
The mean absolute percentage deviation of bond length in the $P4/nmm$ structure is 0
but in the $C2/m$ structure is 10.5\%.
Therefore, we believe that the factors such as
a higher coordination number, shorter bond length
and a more symmetrical environment of the hydrogen atoms,
should be useful to obtain a high $T_c$ superconductor in the H-rich compounds.
Actually similar phenomena can also be found
in the hydrogen sulfide system \cite{Duan2014}.

\begin{figure}[hpt]
\begin{center}
\includegraphics[width=0.45\textwidth]{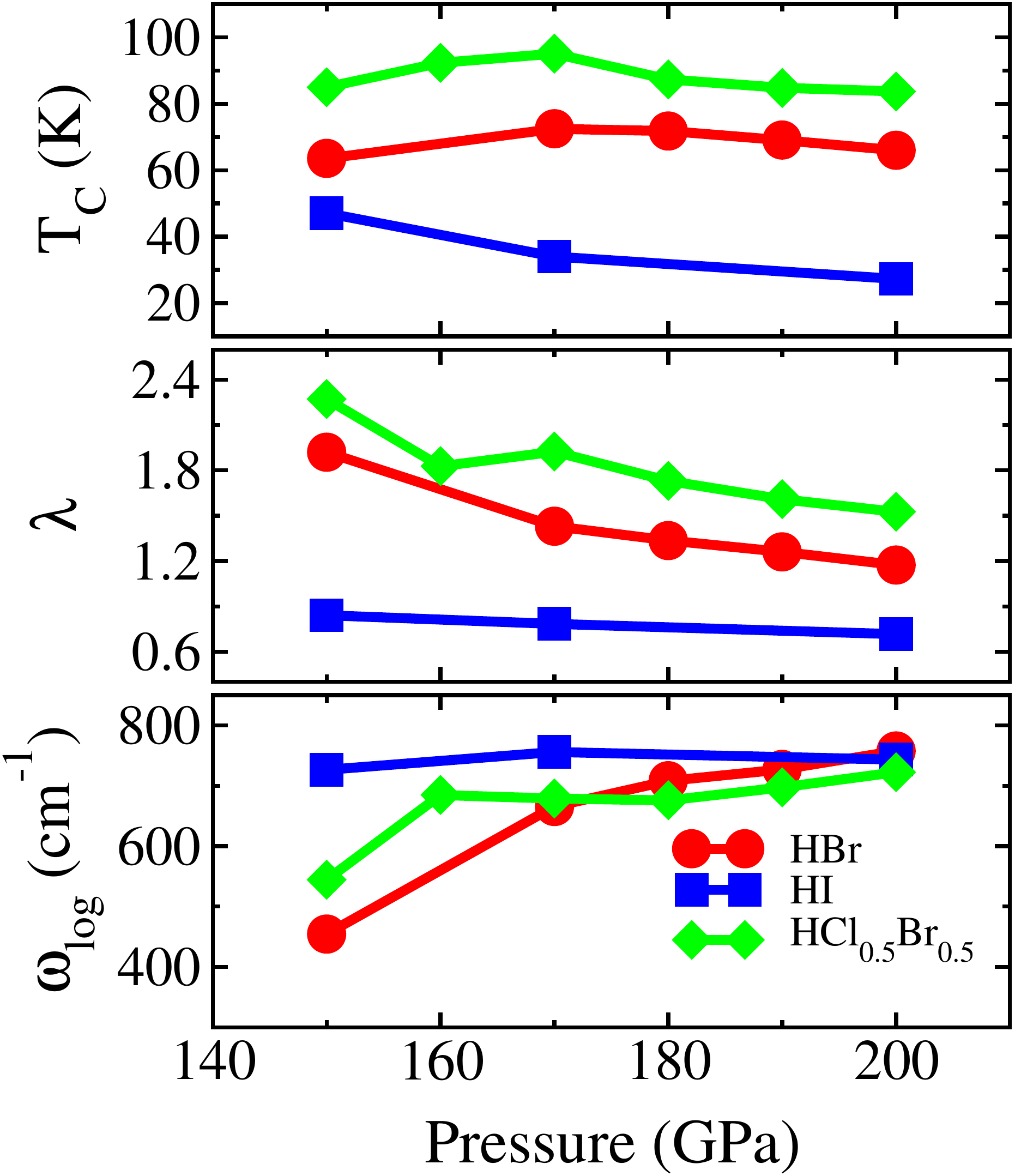}
\caption{\textcolor{blue}{(Color online)}
$T_c$ (red solid line),
the integral EPC parameter $\lambda$ (blue solid line),
and the logarithmically averaged phonon frequency $\omega_{log}$
of $P4mm$-HCl$_{0.5}$Br$_{0.5}$,
$P4/nmm$-HBr, and $P4/nmm$-HI versus pressure.}
\label{fig:$T_c$-pressure}
\end{center}
\end{figure}

\subsection{$T_c$ vs. pressure}

It is obvious that pressure is an effective method to 
increase the density of materials, 
and then to reduce the bond length, get a higher coordination number, 
and symmetrize the crystal structures. 
Therefore, high pressure plays a significant role 
in the research of superconductivity \cite{Gao1994, Duan2014, Zhou2016, Zhou2017}. 
We optimize the $T_c$ of halogen hydrides
by considering pressure. 
The main results are summarized in Fig.\ \ref{fig:$T_c$-pressure}. 
The integral EPC constant $\lambda$ decreases 
as the pressure rises and the logarithmically averaged phonon $\omega_{log}$
 has the opposite trend in general. 
As both of these two parameters have influence 
on superconducting properties, 
the resulting $T_c$ values of halogen hydrides have a complicated evolution with pressure.
 For HCl$_{0.5}$Br$_{0.5}$ and HBr, 
$T_c$ rises and reaches its maximum at around 170 GPa,
 where it falls down afterwards. 
However, in the case of HI, $T_c$ decreases monotonously as the pressure increases. 
The estimated high values of $T_c$ for halogen hydrides are summarized in 
Table.\ \ref{table:$T_c$}.
 The $T_c$ of HCl$_{0.5}$Br$_{0.5}$ in $P4mm$ structure 
is found to be as high as 95 K, 
which is higher than that of all the hydrogen-rich halogen hydrides studied before. 

\section{Conclusions}

In this paper, we present a comprehensive analysis of
the ground state of HBr at pressures ranging from 150 to 200 GPa
based on first-principles calculations.
We find that ZPE has a larger effect on the high-symmetry structure
than the low symmetry ones.
On account of the effect of ZPE,
the ground state of HBr should be the 2D netlike $P4/nmm$ structure,
instead of the chainlike $C2/m$ phase.

Using $ab$ $initio$ perturbative linear response calculations,
we predict that $T_c$ of the $P4/nmm$ phase HBr can reach around 73 K at 170 GPa.
By substituting half of the bromine atoms in the $P4/nmm$ phase with chlorine atoms,
the $T_c$ of resulted $P4mm$ HCl$_{0.5}$Br$_{0.5}$ can reach around 95 K at 170 GPa,
which is the highest record among all the known hydrogen halides so far.
The $P4/nmm$ and $C2/m$ phase are closely related to each other
with additional bonds and sliding transformation.
The energies of these two structures are very close
but their estimated superconducting transition temperature is very different,
specifically, from 73 K for the $P4/nmm$ HBr to almost non-superconducting for the $C2/m$ structure.
The big difference between them is that the hydrogen atoms in $P4/nmm$
are tetrahedral four-coordinated,
while they are two-coordinated in the $C2/m$ structure,
and the HBr bond length in $P4/nmm$ is shorter than that in the $C2/m$ structure.
Our results suggest that in addition to lower atomic mass,
larger coordination number, shorter bonds
and more restricted symmetrical environment for the light atoms
play significant roles in electron-phonon mediated superconductivity.
These factors provide guidance in seeking good superconducting materials
in the future, which are possibly achievable
by pressurization or atomic substitution.

\section{ACKNOWLEDGEMENTS}

We are grateful for the financial support from  
the National Key R\&D program of China (Grant No. 2016YFA0300404), 
the National Key projects for Basic Research in China (Grant No. 2015CB921202)
the National Natural Science Foundation of China (Grant Nos: 51372112 and 11574133), 
NSF Jiangsu province (No. BK20150012), 
the Science Challenge Project (No. TZ2016001),
the Fundamental Research Funds for the Central Universities (No. 020414380068/1-1) 
and Special Program for Applied Research on 
Super Computation of the NSFC-Guangdong Joint Fund (the 2nd phase). 
Part of the calculations was performed on the supercomputer 
in the HPCC of Nanjing University and ``Tianhe-2'' at NSCC-Guangzhou.

\section{APPENDIX}

\begin{table}[hpt]
\centering
  \caption{
The calculated decomposition enthalpy
(defined as $\Delta$H =
H$_{HCl_{0.5}Br_{0.5}}$-\(\frac12\)H$_{H_2}$-\(\frac14\)H$_{Cl_2}$-\(\frac14\)H$_{Br_2}$)
for the $P4mm$ phase
under high pressure,
based on the PBE-D2 functional.
}
 \begin{tabular}[t]{m{3.2cm}m{1.5cm}m{1.5cm}p{1.5cm}}
        \hline\hline
        Pressure (GPa)& 150 & 170 & 200 \\
        \hline
        $\Delta$H (meV) & -157.26 & -169.44 & -189.06  \\
        \hline\hline
  \end{tabular}
\label{table:HClBrenthalpy}
\end{table}

\begin{table}[hpt]
\centering
  \caption{
Detailed structure information 
of the halogen hydrides 
at selected pressures.
}
 \begin{tabular}[t]{m{1.5cm}m{1.3cm}m{1.8cm}L{3.8cm}}
	\hline\hline
	Phase & Pressure  & Lattice & Atomic coordinates \\
	      &  (GPa)	 & parameters (\AA) &  (fractional)\\
	\hline
	$P4/nmm$  & 170 & a = 3.190 & H(2b)\quad0.500 0.500 0.500\\
	Hbr	     &     & c = 2.636 & Br(2c)\quad0.500 0.000 0.795\\
	\hline
	$P4/nmm$   & 150 & a = 3.488 & H(2b)\quad0.500 0.500 0.500\\
	  HI       &     & c = 2.823 & I(2c)\quad 0.500 0.000 0.745\\
	\hline
	$P4mm$       & 170 & a = 3.074 & H(2c)\quad0.000 0.500 0.486\\
	HCl$_{0.5}$Br$_{0.5}$ & & c = 2.618 & Cl(1a)\quad0.000 0.000 0.811\\
	                      & &           & Br(1b)\quad0.500 0.500 0.217\\
	\hline\hline
   \end{tabular}
\label{table:coordinate}
\end{table}

\begin{table}[hpt]
\centering
  \caption{
EPC parameter ($\lambda$),
logarithmic average of phonon frequencies ($\omega_{log}$), and
estimated superconducting critical temperature ($T_c$)
with the Coulomb potential ($\mu^*$) of $0.1$
for $P4mm$-HCl$_{0.5}$Br$_{0.5}$, $P4/nmm$-HBr, and $P4/nmm$-HI
under high pressures.
}
 \begin{tabular}[t]{m{2cm}m{1.5cm}m{1.3cm}m{1.5cm}m{1.2cm}}
        \hline\hline
        Phase  & Pressure (GPa)& $\lambda$ & $\omega_{log}$ (cm$^{-1}$) & $T_c$ (K)  \\
        \hline
        $P4/nmm$ HBr & 170 & 1.43 & 665.7 & 72.5  \\
        \hline
        $P4/nmm$  HI  & 150 & 0.84 & 726.3 & 37.6  \\
        \hline
        $P4mm$ HCl$_{0.5}$Br$_{0.5}$ & 170 & 1.93 & 678.9 & 95.1  \\
        \hline\hline
  \end{tabular}
\label{table:$T_c$}
\end{table}


\end{document}